\documentclass[12pt]{article}
\usepackage{amsfonts}

\usepackage{graphicx}
\usepackage{amsmath}
\usepackage{float}

\textwidth=6.5in \hoffset=-.75in \voffset=-1in \numberwithin{equation}{section}
\numberwithin{figure}{section}

\begin{document}

\begin{titlepage}
\vspace{1cm}
\begin{center}
{\Large \bf {Kerr-Bolt Spacetimes and Kerr/CFT Correspondence}}\\
\end{center}
\vspace{2cm}
\begin{center}
A. M. Ghezelbash{ \footnote{ E-Mail: masoud.ghezelbash@usask.ca}}
\\
Department of Physics and Engineering Physics, \\ University of Saskatchewan, \\
Saskatoon, Saskatchewan S7N 5E2, Canada\\
\vspace{1cm}
\end{center}

\begin{abstract}
We investigate the recently proposed Kerr/CFT correspondence in the context of rotating spacetimes with a NUT twist. The Kerr/CFT correspondence states that the near-horizon states of an extremal four (or higher) dimensional black hole could be identified with a certain chiral conformal field theory. The corresponding Virasoro algebra is generated with a class of diffeomorphism which preserves an appropriate boundary condition on the near-horizon geometry. We try to understand the analog of singularities in the context of dual chiral CFT. Explicitly, we use Kerr/CFT correspondence to show that if we initially do not remove the singularities from the spacetimes, in the dual chiral CFT we can detect the presence of singularities in the bulk of spacetime. 

\end{abstract}
\end{titlepage}\onecolumn 
\bigskip 

\section{Introduction}
For a long time, black holes have been an interesting theoretical system to understand the nature of quantum gravity. Despite a lot of efforts to explain and reproduce the Bekenstein-Hawking entropy, the theory of black hole entropy is not complete. 

Recently, in the context of proposed Kerr/CFT correspondence \cite{stro}, (see also \cite{Solo} and \cite{Bar}) the microscopic entropy of four-dimensional extremal Kerr black hole is calculated by studying the dual chiral conformal field theory associated with the diffeomorphisms of near horizon geometry of the Kerr black hole. These diffeomorphisms preserve an appropriate boundary condition at the infinity. One important feature of this correspondence is that it doesn't rely on supersymmetry and string theory unlike the well known AdS/CFT correspondence. \cite{ads1,ads2,ads3,ads4,ads5,ads6}.

The Kerr/CFT correspondence has been used in \cite{LU} and \cite{cvetic} to find the entropy of dual CFT for 
four and higher dimensional Kerr black holes in AdS spacetimes and gauged supergravity as well as five-dimensional BMPV black holes in  \cite{ISO}. Moreover the correspondence has been used in string Theory D1-D5-P and BMPV black holes in \cite{Aze} 
and in the five dimensional Kerr black hole in G\"{o}del universe \cite{G1}.
The continuous approach to the extremal Kerr black hole is essential in the proposed correspondence. For example, in the case of Reissner-Nordstrom black hole the approach to extremality is not continuous \cite{car}. The rotating bubbles, Kerr-Newman black holes in (A)dS spacetimes and rotating NS5 branes have been considered in \cite{Aze2}, \cite{Stro2} and \cite{Naka}. The four-dimensional Kerr-Sen black hole has been considered in \cite{GH2}.

Inspired with these works, in this article, we consider the class of Kerr-Bolt spacetimes. The Kerr-Bolt spacetimes are exact solutions to the four-dimensional gravity. 
The spacetimes with NUT twist have been studied extensively in regard to their conserved charges, maximal mass conjecture and D-bound in \cite{TN1}.
These spactimes have conical and Dirac-Misner singularities that should be removed by identifications of coordinates in the metric.

We apply the Kerr/CFT correspondence to the extremal Kerr-Bolt spacetimes (these extremal spacetime exist if we don't remove the singularities from Kerr-Bolt spacetimes) and conclude that in the context of Kerr/CFT correspondence, the results from CFT side can show and detect the presence of singularities in the bulk. The results of this paper are all in favor of Kerr/CFT correspondence.

The outline of this paper is as follows. In section \ref{sec:5Dreview},
we first review briefly the rotating spacetimes with the NUT charge and show that the only consistent rotating spacetimes with the NUT charge are Kerr-Bolt spacetimes where the fixed point set of the Killing vector is two dimensional sphere. In section \ref{sec:near}, we find the near-horizon geometry of extremal spacetimes by
using a special coordinate transformations. In section \ref{sec:sol},
we calculate the central charge and microscopic entropy of the extremal Kerr-Bolt spactimes. We then study the fist law of thermodynamics in CFT and find that the first law of thermodynamics cannot be satisfied except if the spacetime is free of any NUT charge. This in turn, indicates that the NUT charges induce singularities in the spacetime. In section \ref{sec:non-ex}, we review how we should define a positive definite and consistent expression with the first law, for the entropy of non-extremal Kerr-Bolt spacetimes. We conclude in section \ref{sec:con} with a summary of our results.

\section{Kerr-Bolt Spacetimes}

\label{sec:5Dreview}

In this section, we give a brief review of the Kerr-Bolt spacetime.
The Lorentzian geometry of Kerr spacetimes with NUT charge and nonzero
cosmological constant is given by the line element%
\begin{equation}
ds^{2}=-\frac{\Delta _{L}(r)}{\Xi _{L}^{2}\rho _{L}^{2}}[dt+(2n\cos \theta
-a\sin ^{2}\theta )d\varphi ]^{2}+\frac{\Theta _{L}(\theta )\sin ^{2}\theta }{%
\Xi _{L}^{2}\rho _{L}^{2}}[adt-(r^{2}+n^{2}+a^{2})d\varphi ]^{2}+\frac{\rho
_{L}^{2}dr^{2}}{\Delta _{L}(r)}+\frac{\rho _{L}^{2}d\theta ^{2}}{\Theta
_{L}(\theta )}  \label{LKBdS}
\end{equation}%
where%
\begin{equation}
\begin{array}{c}
\rho _{L}^{2}=r^{2}+(n+a\cos \theta )^{2} \\ 
\Delta _{L}(r)=-\frac{r^{2}(r^{2}+6n^{2}+a^{2})}{\ell ^{2}}+r^{2}-2mr-\frac{%
(3n^{2}-\ell ^{2})(a^{2}-n^{2})}{\ell ^{2}} \\ 
\Theta _{L}(\theta )=1+\frac{a\cos \theta (4n+a\cos \theta )}{\ell ^{2}} \\ 
\Xi _{L}=1+\frac{a^{2}}{\ell ^{2}}%
\end{array}
\label{TNKdSLfuncs}
\end{equation}%
which are exact solutions of the Einstein equations.
The ``event horizons'' of
the spacetime are given by the singularities of the metric function which
are the real roots of $\Delta _{L}(r)=0$. These are determined by the
solutions of the equation%
\begin{equation}
  r_{+}^{4}-r_{+}^{2}(\ell ^{2}-6n^{2}-a^{2})
  +2m\ell^{2}r_{+}+(3n^{2}-\ell^{2})(a^{2}-n^{2})=0  
 \label{TNKDShorizons}
\end{equation}

The ``Euclidean section'' for this class of metrics is given by 
\begin{equation}
ds^{2}=\frac{\Delta _{E}(r)}{\Xi _{E}^{2}\rho _{E}^{2}}[dt-(2n\cos \theta
-a\sin ^{2}\theta )d\varphi ]^{2}+\frac{\Theta _{E}(\theta )\sin ^{2}\theta }{%
\Xi _{E}^{2}\rho _{E}^{2}}[adt-(r^{2}-n^{2}-a^{2})d\varphi ]^{2}+\frac{\rho
_{E}^{2}dr^{2}}{\Delta _{E}(r)}+\frac{\rho _{E}^{2}d\theta ^{2}}{\Theta
_{E}(\theta )} \label{EKBdS}
\end{equation}%
where we have analytically continued the $t$ coordinate, the nut charge and
the rotation parameter to imaginary values, yielding%
\begin{equation}
\begin{array}{c}
\rho _{E}^{2}=r^{2}-(n+a\cos \theta )^{2} \\ 
\Delta _{E}(r)=-\frac{r^{2}(r^{2}-6n^{2}-a^{2})}{\ell ^{2}}+r^{2}-2mr-\frac{%
(3n^{2}+\ell ^{2})(a^{2}-n^{2})}{\ell ^{2}} \\ 
\Theta _{E}(\theta )=1-\frac{a\cos \theta (4n+a\cos \theta )}{\ell ^{2}} \\ 
\Xi _{E}=1-\frac{a^{2}}{\ell ^{2}}%
\end{array}%
\end{equation}

The Euclidean section exists only for values of $r$ such that the function $%
\Delta _{E}(r)$ is positive valued. The horizons are located at the zeros of 
$\Delta _{E}(r)$, which we shall denote by $r=r_{0}$. Moreover, the range of 
$\theta $ depends strongly on the values of the NUT charge $n$, the
rotational parameter $a$ and the cosmological constant $\Lambda =3/\ell ^{2}$%
, taken here to be positive (for the solution with $\Lambda <0$, replace $%
\ell ^{2}$ with $-\ell ^{2}$ in the preceding equations). The angular
velocity of the horizon in the Lorentzian geometry is given by%
\begin{equation}
\Omega _H =\left. -\frac{g_{t\varphi }}{g_{\varphi \varphi }}\right| _{r=r_{0}}=%
\frac{a}{r_{0}^{2}+n^{2}+a^{2}}  \label{surface}
\end{equation}%
and so the surface gravity of the cosmological horizon can be calculated to
give%
\begin{equation}
\kappa =\frac{1}{2(r_{0}^{2}+n^{2}+a^{2})\Xi_L }\left. \frac{d\Delta _L }{d r }
\right| _{r=r_{0}}
\end{equation}%
where the Killing vector 
$\chi ^{\mu }=\zeta ^{\mu }+\Omega_H \psi^{\mu}$
is normal to the horizon surface $r=r_{0}$.

We first note that there are no pure NUT solutions for nonzero values of $a$. 
We demonstrate this as follows.\footnote
{It also follows purely topologically from the fact that the surfaces
 of constant $r$ are not homeomorphic to $S^3$.}
Since $\psi ^{\mu }$\ is a Killing vector,
for any constant $\varphi $ -slice near the horizon, additional conical
singularities will be introduced in the $(t,r)$\ section unless the period
of $t$\ is $\Delta t=\frac{2\pi }{\left| \kappa \right| }$. Furthermore,
there are string-like singularities along the $\theta =0$ and $\theta =\pi $
axes for general values of the parameters. \ These can be removed by making
distinct shifts of the coordinate $t$ in the $\varphi $ direction near each of
these locations. These must be geometrically compatible 
\cite{GHKB},
yielding the requirement that the period of $t$ should be
$\Delta t=4n\Delta \phi=\frac{16\pi n}{q_++q_-}$. Demanding the absence of both
conical and Dirac-Misner singularities, we get the relation 
\begin{equation}
  \frac{k}{\kappa}=\frac{8n}{q_++q_-} 
  \label{betaeq8Pin}
\end{equation}%
where $k$ is any non-zero positive integer. Demanding the existence of a
pure NUT solution at $r=r_{0}$ is equivalent to the requirement that that
the area of the surface of the fixed point set of the Killing vector $%
\partial /\partial t,$%
\begin{equation}
 A=\frac{2\pi}{\Xi_{E}} 
   \{ \alpha \sqrt{\Delta_E(r_{0})} + 2(r_{0}^{2} - n^2-a^2) \}  
  \label{surfacefixed}
\end{equation}%
vanish, 
where $\alpha =\int_{0}^{\pi }\frac{2n\cos \theta -A\sin ^{2}\theta}
{\sqrt{\Theta _{E}(\theta )}}d\theta $. 
In other words, this surface is of
zero dimension. This can only occur special values of the mass parameter.
However if we select for this parameter we find an inconsistency with the
relation (\ref{betaeq8Pin}), which must hold for the spacetime with NUT
charge.

Hence we conclude that the only spacetimes with NUT charge and rotation is
Taub-Bolt-Kerr-(A)dS spacetimes (or simply Kerr-Bolt-(A)dS), where the term ``bolt'' refers to the fact
that the dimensionality of the fixed point set of $\partial /\partial t$
is two.

In this article, we consider Kerr/CFT correspondence in the asymptotically flat case and leave the cases with cosmological constant for further investigations.

\section{Near-Horizon Geometry}
\label{sec:near}

In this section, we study the near-horizon geometry of Kerr-Bolt spacetimes without any constraint as (\ref{betaeq8Pin}).
The Kerr-Bolt spacetime is given by the line element
\begin{equation}
ds^{2}=-\frac{\Delta (\tilde r)}{\rho ^{2}}[dt+(2n\cos \theta
-a\sin ^{2}\theta )d\tilde \phi ]^{2}+\frac{\sin ^{2}\theta }{%
\rho^{2}}[ad \tilde t-(\tilde r^{2}+n^{2}+a^{2})d\tilde \phi ]^{2}+\frac{\rho
dr^{2}}{\Delta (\tilde r)}+\rho ^{2}d\theta ^{2}
\label{BH}
\end{equation}
where
\begin{eqnarray}
\rho ^{2}=\tilde r ^{2}+(n+a\cos \theta )^{2} \\ 
\Delta (r)=\tilde r ^{2}-2m\tilde r+a^{2}-n^{2}
\end{eqnarray}
The event horzon is located at $r_0=m+\sqrt{m^2-a^2+n^2}$ with angular velocity (\ref{surface}). The Hawking temperature is equal to
\begin{equation}
T_H=\frac{\sqrt{m^2-a^2+n^2}}{2\pi(r_0^2+n^2+a^2)}\label{Hawking}
\end{equation}
To find the near-horizon limit of the extreme black hole, we change the coordinates by the following transformations 
\begin{eqnarray}
\tilde r&=&m(1+\frac{\lambda}{y})\\
\tilde t&=& \frac{2a^2}{r_0\lambda}t \\
\tilde \phi &=& \phi + \frac{a}{r_0}t/\lambda
\end{eqnarray}
where the scaling parameter $\lambda$ approaches zero. The metric (\ref{BH}) changes then to the near-horizon metric given by
\begin{eqnarray}
ds^2&=&\{a^2(1+\cos^2(\theta))+2na\cos(\theta)\}
\{\frac{-dt^2+dy^2}{y^2}+d\theta^2\}\nonumber\\
&+&\frac{4a\sin^2(\theta)}{a(1+\cos^2(\theta))+2n\cos(\theta)}(ad\phi+\sqrt{a^2-n^2}\frac{dt}{y})^2
\label{BH2}
\end{eqnarray}
The metric definitely is not asymptotically flat. To cover the whole near-horizon geometry, we use the global coordinates
\begin{eqnarray}
y&=&\frac{1}{\cos\tau\sqrt{1+r^2}+r}\\
t&=&y\sin\tau\sqrt{1+r^2}\\
\phi&=&\frac{\sqrt{a^2-n^2}}{a}\{\varphi +\ln (\frac{\cos\tau+r\sin\tau}{1+\sin\tau\sqrt{1+r^2}})\}
\end{eqnarray}

The global near-horizon metric is
\begin{eqnarray}
ds^2&=&\{a^2(1+\cos^2(\theta))+2na\cos(\theta)\}
\{-(1+r^2)d\tau^2+\frac{dr^2}{1+r^2}+d\theta^2+\nonumber\\
&+&\frac{4\sin^2(\theta)}{\{a(1+\cos^2(\theta))+2n\cos(\theta)\}^2}(a^2-n^2)      (d\varphi+rd\tau)^2\}\label{global}
\end{eqnarray}
In the case of vanishing NUT charge, the metric becomes the near-horizon geometry of the Kerr solution, as in \cite{stro,Bardeen}. For a fixed $\theta$, the near-horizon geometry is a quotient of warped AdS$_3$ which the quotient arises from identification of $\varphi$ coordinate. The isometry group of the geometry is $SL(2,R)\times U(1)$, where $U(1)$ is generated by the Killing vector $-\partial _\varphi$ and $SL(2,R)$ is generated by three Killing vectors,
\begin{eqnarray}
J_1&=&2\sin\tau\frac{r}{\sqrt{1+r^2}}\partial_\tau-2\cos\tau\sqrt{1+r^2}\partial_r+\frac{2\sin\tau}{\sqrt{1+r^2}}\partial_
\varphi\\
J_2&=&-2\cos\tau\frac{r}{\sqrt{1+r^2}}\partial_\tau-2\sin\tau\sqrt{1+r^2}\partial_r-\frac{2\cos\tau}{\sqrt{1+r^2}}\partial_
\varphi\\
J_3&=&2\partial_\tau
\end{eqnarray}

\section{Microscopic Entropy in CFT and Failure of First Law}
\label{sec:sol}

Choosing the proper boundary condition for the near-horizon metric as the same as one in \cite{stro}, it can be shown that the near-horizon metric has a class of commuting diffeomorphisms labeled by $p=0,\pm 1,\pm 2, \cdots$
\begin{equation}
\zeta_p=-e^{-ip\varphi}(\partial_\varphi+ipr\partial_r)
\end{equation}
This diffeomorphism generates a Virasoro algebra without any central charge
\begin{equation}
[\zeta _p,\zeta _q]=-i(p-q)\zeta_{p+q}
\end{equation}

The generator of $\zeta_p$ is the conserved charge $Q_\zeta[g]$ between neighboring geometries $g$ (which is the near-horizon metric (\ref{global})) and $h$ (which is the deviation from the near-horizon metric (\ref{global})) and is given by
\begin{equation}
Q_{\zeta_p}[g]=\frac{1}{8\pi}\int_{\partial\Sigma} k_{\zeta _p}[h,g] 
\end{equation}
In above equation, $\partial \Sigma $ is the boundary of a spatial slice and two form
$k_{\zeta}$ is
\begin{eqnarray}
k_\zeta[h,g]&=&-\frac{1}{4}\epsilon_{\mu\nu\rho\sigma}\{\zeta^\sigma\nabla^\rho h-\zeta^\sigma\nabla_\lambda h^{\rho\lambda}+\zeta_\lambda\nabla^\sigma h^{\rho\lambda}+\frac{1}{2}h\nabla ^{\sigma}\zeta^\rho\nonumber\\
&-&h^{\sigma\lambda}\nabla_\lambda\zeta^\rho+\frac{1}{2}h^{\lambda\sigma}(\nabla ^
\rho\zeta_\lambda+\nabla_\lambda\zeta^\rho)\}dx^\mu \wedge dx^\nu
\end{eqnarray}
The Dirac brackets of two conserved charges $Q_{\zeta_p}$ and $Q_{\zeta_q}$ yields $Q_{[\zeta_p,\zeta_q]}$ and a central term which is equal to $\frac{1}{8\pi}\int_{\partial\Sigma}k_{\zeta _p}[{\cal L}_{\zeta_q}g,g]$, where 
${\cal L}_{\zeta_q}g$ is the Lie derivative of 
the near-horizon metric (\ref{global}). Replacing the Dirac brackets by commutators and straightforward calculation of the central term yields a Virasoro algebra with the central charge
\begin{equation}
c=12a\sqrt{a^2-n^2}\label{centralcharge}
\end{equation}
for the dual chiral CFT corresponding to Kerr-bolt spacetime (\ref{BH2}).
To find the entropy of dual chiral CFT, we need to find Frolov-Thorne temperature \cite{Fro}.
A straightforward calculation shows the right temperature $T_R=0$ and the left temperature
\begin{equation}
T_{L}=\frac{a}{2\pi\sqrt{a^2-n^2}}\label{TL}
\end{equation}
hence the Frolov-Thorne temperature is $T_{FT}=T_L$. Finally, we get the microscopic entropy in dual chiral CFT by using the Cardy relation
\begin{equation}
S=\frac{\pi^2}{3}cT_{FT}=2\pi a^2=2\pi (m^2+n^2)\label{mic}
\end{equation}
In the special case of $n=0$, this result is exactly in agreement with the macroscopic Bekenstein-Hawking entropy of Kerr black hole \cite{stro}. On the other hand for non-zero NUT charge $n$, the microscopic result (\ref{mic}) does not satisfy the microscopic first law of thermodynamics. In fact, we show the first law of thermodynamics gives for the microscopic temperature something quite different from Frolov-Thorne temperature given by (\ref{TL}). To see this, we consider the first law as
\begin{equation}
T dS=dM-\Omega _H dJ
\end{equation}
Here we consider the entropy as a function of $M=m$ and $J=ma$ where we choose $q_+=q_-=1$ for simplicity. (see section (\ref{sec:non-ex}) for details). By introducing a parameter $\delta=m-\sqrt{a^2-n^2}$ which measures non-extremality, the entropy can be considered as a function of $\delta$ and $J$. So the first law can be rewritten as 
\begin{equation}
dS=\beta _Hd\delta+\beta dJ
\end{equation}
where $\beta=(\frac{\partial S}{\partial J})_{\text{fixed \,} \delta}$. A straightforward calculation shows 
\begin{equation}
\beta=\beta_H\{\frac{2J}{\sqrt{n^4+4J^2}\sqrt{-2n^2+2\sqrt{n^4+4J^2}}}-\frac{1}{\sqrt{2n^2+2\sqrt{n^4+4J^2}}}\}
\label{beta}
\end{equation}
where $\beta_H=1/T_H$. This equation gives the temperature of chiral CFT as $T=0$, not in agreement with (\ref{TL}). We note that the only consistent result from equation (\ref{beta}) with Frolov-Thorne temperature (\ref{TL}), is in the special case of $n=0$. In this case (\ref{beta}) reproduces exactly $T=\frac{1}{2\pi}$ in agreement with the temperature of dual CFT to Kerr black hole. So we conclude applying Kerr/CFT correspondence to the metric (\ref{LKBdS}) without removing its conical and Misner-Dirac singularities (given by equation (\ref{betaeq8Pin})), leads to violation of the microscopic first law in CFT side. 

\section{Entropy of non-extremal Kerr-Bolt Spacetimes}
\label{sec:non-ex}

In this section, we discuss briefly about the entropy of non-extremal Kerr-Bolt spacetimes in both Euclidean and Lorentzian signatures. Although the entropy of non-extremal black holes was discussed in literature \cite{Mann}, but it can be shown that the results don't satisfy the first law of thermodynamics. The reason for this (as we see later in this section) is that the singularities on the north and south poles as well as on horizon should be removed consistently. This leads to introducing three vectors that generate a lattice and they should be minimal and commensurable (see equations (\ref{etaplus}), 
(\ref{etaminus}), (\ref{xidef}), (\ref{lattice})).

All the instantons we will consider will be derived from Kerr-Bolt line
element by identifications. As we will see, these identifications will not
only twist the imaginary-time direction near infinity, but also modify even
the topology of the ``spatial'' two-sphere at infinity, somewhat in the
manner of the so called ``Asymptotically Locally Euclidean instantons''.
That is, the topology at infinity will (unfortunately) not necessarily be
that of a circle bundle over $S^{2}$.

The line element is singular along the polar axes $\theta =0,\pi $ and at
the ``horizon'' $r=r_{0}$. The ``string'' singularities at the poles have,
as is well known, a more complicated structure than that belonging to the
familiar spherical coordinates for $S^{2}$, and they cannot be removed
merely by the imposition of periodicity in azimuth $\varphi $. Rather two
separate identifications are required to remove the two ``strings''.
Moreover, one needs a further identification to remove the singularity at
the horizon, and all of the identifications must be compatible. This
compatibility condition is crucial, but it seems to have been overlooked in
the literature on these metrics. To analyze it, we may refer everything to a
lattice in the time-azimuth plane.

The $\ $vectors $\partial /\partial {t}$ and $\partial /\partial {\varphi }$
are both Killing vectors so this makes it possible to quotient our metric by
any linear combination of the two (with constant coefficients). More
generally, we can quotient by any {\it lattice} $\cal{L} $ of vectors which is
closed under vector sum and difference. The resulting spacetime is entirely
determined by $\cal{L} $. Let us consider how $\cal{L} $ must be chosen in order to
remove the two types of singularity we have to deal with.

To take advantage of the scale invariance of our family of metrics, we will
introduce in place of $t$ the dimensionless coordinate 
\begin{equation}
\psi = t / 2n \ ,
\end{equation}
and the corresponding Killing vector 
\begin{equation}
\partial / \partial\psi = 2n \; \partial / \partial{t} \ .
\end{equation}

Now look at the neighborhood of the north polar axis, $\theta =0$. At the
pole, the line-element acquires a degenerate direction. In order to
compensate, we must quotient by a vector $\eta _{+}$ that is parallel to
this degenerate direction and whose length is chosen so that the quotient
metric will not exhibit a conical singularity. That is, the length of $\eta
_{+}$ at a proper distance $\varepsilon $ from the north pole, must be $2\pi
\varepsilon $ to first order in $\varepsilon $. Some algebra shows that the
required vector is \cite{GHKB}
\begin{equation}
\eta _{+}=2\pi \,(\partial /\partial \psi +\partial /\partial \varphi )\ .
\label{etaplus}
\end{equation}%
This therefore is one of the vectors of $\cal{L} $, but we can say more. It is
also necessary that no ``submultiple'' of $\eta _{+}$ belong to $\cal{L} $;
otherwise we would really be quotienting by a smaller vector. That is, $\eta
_{+}$ must be a ``minimal'' element of the lattice. (By minimal we mean that 
$\lambda \eta _{+}\notin \cal{L} $ for $0<\lambda <1$.) The same analysis at the
south pole furnishes a second minimal lattice vector, 
\begin{equation}
\eta _{-}=2\pi (\partial /\partial \psi -\partial /\partial \varphi )\ .
\label{etaminus}
\end{equation}

The singularity at $r=r_0$ can be analyzed similarly. Although the algebra
is a bit messier, it proceeds along the same lines and reveals a third
minimal lattice vector, 
\begin{equation}
\xi = \beta_H (\partial /\partial t + \Omega_H \partial /\partial \varphi )
\ ,  \label{xidef}
\end{equation}
where $\beta_H$ and $\Omega_H$ are given by
\begin{equation}
\beta_H=\frac{4\pi(r_0^2-n^2-a^2)r_0}{r_0^2-n^2+a^2}
\end{equation}
and
\begin{equation}
\Omega_H=\frac{a}{r_0^2-n^2-a^2}
\end{equation}

Finally, there are the special points where the polar axes meet the horizon.
When the angular velocity vanishes, no extra conditions on $\cal{L} $ are imposed
by regularity at these special points; by analyticity there is no reason to
suppose that extra conditions arise when rotation is ``turned on''.

What does impose a crucial extra condition, however, is the requirement that
our quotient spacetime be a manifold. If the lattice $\cal{L} $ were dense in the 
$\varphi $-$t$-plane, this obviously would not be the case, as the quotient $%
{\mathbb R}^{2}/\cal{L} $ would be pathologically non-Hausdorff. In order to
preclude this, one must arrange that three vectors $\eta _{\pm }$ and $\xi $
be {\it commensurate}. (In other words, $\xi $ must be a linear combination
of $\eta _{\pm }$ with rational coefficients.) Bringing in the minimality
constraints as well, one can see that the complete condition is 
\begin{equation}
p\xi =q_{+}\eta _{+}+q_{-}\eta _{-}\ ,  \label{lattice}
\end{equation}%
where $p$, $q_{+}$ and $q_{-}$ are three relatively prime integers: $%
(p,q_{+})=(p,q_{-})=(q_{+},q_{-})=1$.

Notice that these conditions have the exceptional solution, $p=q_{+}=1$, $%
q_{-}=0$. For this solution, $\xi $ coincides with $\eta _{+}$, but in all
other cases (except for the mirror image case $\xi =\eta _{-}$) all three
vectors point in distinct directions. This exceptional solution will also be
exceptional in its thermodynamic properties. We don't consider this
exceptional case in this article.

{}From (\ref{lattice}), we find 
\begin{equation}
\beta _{H}=4\pi n\ {\frac{q_{+}+q_{-}}{p}}\ ,  \label{betaHc}
\end{equation}%
\begin{equation}
\Omega _{H}=\frac{1}{2n}{\frac{q_{+}-q_{-}}{q_{+}+q_{-}}}\ ,  \label{Om}
\end{equation}%
and therefore 
\begin{equation}
\widetilde{\Omega }_{H}=\beta _{H}\Omega _{H}=2\pi {\frac{q_{+}-q_{-}}{p}}\ .
\label{OtH}
\end{equation}

We emphasize that the parameters $p$, $q_{\pm }$ are integers and therefore
not continuously variable. For this reason --- and in sharp contrast to
instantons without NUT charge --- the solutions considered here fall into
disconnected families between which no continuous transition is possible. \
As a result, for a fixed set of $p,q_{+}$ and $q_{-},$ the Euclidean metric
has only one single continuous parameter. We can proceed now and compute the
conserved mass ${\cal M}$ and angular momentum ${\cal J}$, and we get 
\begin{equation}
{\cal M}={\frac{2}{q_{+}+q_{-}}}m\ ,  \label{Mcons}
\end{equation}%
and 
\begin{equation}
{\cal J}={\frac{2}{q_{+}+q_{-}}}am\ .  \label{Jcons}
\end{equation}%
The total Euclidean action, expressed in terms of the rationalized
dimensionless variables $\widehat{\beta }=\beta /4\pi n$, and $\widehat{%
\Omega }=\widetilde{\Omega }/2\pi $, is given by 
\begin{equation}
I_{E}={\frac{4\pi n^{2}}{p}}{\frac{1+\widehat{\beta }^{2}-\widehat{\Omega }%
^{2}}{\widehat{\beta }+\sqrt{(1-\widehat{\Omega }^{2})(\widehat{\beta }^{2}-%
\widehat{\Omega }^{2})}}}\ . \label{IEuc1}
\end{equation}%
The entropy could be obtained from 
\begin{equation}
S=\beta {\cal M}-\widetilde{\Omega }{\cal J}-I_{E},  \label{ss}
\end{equation}%
where we have taken the thermodynamic quantities as the conserved quantities
and analytically continue $\widehat{\Omega }$ to $-i\widehat{\Omega }$ (see
more details in \cite{GHKB}) and is given by

\begin{equation}
S=\frac{4\pi n^{2}}{p\widehat{\beta }}\frac{1+\widehat{\beta }^{2}+\widehat{%
\Omega }^{2}}{\widehat{\beta }+\sqrt{(1+\widehat{\Omega }^{2})(\widehat{%
\beta }^{2}+\widehat{\Omega }^{2})}}(2\widehat{\beta }-\sqrt{\frac{\widehat{%
\beta }^{2}+\widehat{\Omega }^{2}}{1+\widehat{\Omega }^{2}}}).  \label{s}
\end{equation}%
\bigskip The\ entropy (\ref{s}) is \ definitely positive for all allowed
values of $\widehat{\beta }$ and $\widehat{\Omega }.$ In fact one can show
that $0\leq \widehat{\Omega }^{2}\leq 1\leq \widehat{\beta }$ with $\widehat{%
\Omega }=1$ only if $\widehat{\beta }$ $=1$.

\section{Concluding Remarks}

\label{sec:con}
In this paper, we considered the class of Kerr-Bolt spacetimes without removing their singularities and showed that by using recently Kerr/CFT correspondence, the CFT could detect the presence of singularities. We found explicitly the near-horizon metric of the extremal spacetimes by taking the near-horizon procedure. The near-horizon geometry has topology of warped AdS$_3$. By choosing the proper boundary condition, we can find the diffeomorphism that generates a Virasoro algebra without any central charge. The generator of diffeomorphism which is a conserved charge can be used to construct an algebra under Dirac brackets. This algebra is the same as diffeomorphism algebra but just with an extra central term. The central charge of the Virasoro algebra together with Frolov-Thorne temperature enable us to find the microscopic entropy of the extremal spactime in dual chiral CFT. We showed the microscopic entropy doesn't satisfy the microscopic first law of thermodynamics unless the NUT charge of spacetime vanishes. This in turn show
the metric (\ref{LKBdS}) without removing its conical and Misner-Dirac singularities (given by equation (\ref{betaeq8Pin})), leads to violation of the microscopic first law in CFT side. Our work provides further supportive evidence in favor of a Kerr/CFT correspondence.

\bigskip
{\Large Acknowledgments}

This work was supported by the Natural Sciences and Engineering Research
Council of Canada.

\end{document}